\documentclass[aps,prd,reprint,amsmath,amssymb,superscriptaddress,nofootinbib,nolongbibliography,floatfix]{revtex4-2}
\usepackage{graphicx}
\usepackage[utf8]{inputenc}
\usepackage{xcolor}
\usepackage[unicode=true,colorlinks=true,linkcolor=blue,anchorcolor=blue,urlcolor=blue,citecolor=blue,breaklinks=true,hyperindex]{hyperref}

\newcommand{\ii}{\mathrm{i}}
\newcommand{\pararrow}{\mathord{\buildrel{\lower3pt\hbox{$\scriptscriptstyle\leftrightarrow$}}\over {\partial}}} % partial leftrightarrow operation
\newcommand{\pararrowk}[1]{\mathord{\buildrel{\lower3pt\hbox{$\scriptscriptstyle\leftrightarrow$}}\over {\partial}\hspace*{-0.18em}{}^#1}\hspace*{-0.18em} } % partial

\newcommand{\qfnu}{\affiliation{College of Physics and Engineering, Qufu Normal University, Qufu 273165, China}}

\newcommand{\itp}{\affiliation{CAS Key Laboratory of Theoretical Physics, Institute of Theoretical Physics,\\ Chinese Academy of Sciences, Beijing 100190, China}}

\newcommand{\hnnu}{\affiliation{Institute of Particle and Nuclear Physics, Henan Normal University, Xinxiang 453007, China}}

\newcommand{\ucas}{\affiliation{School of Physical Sciences, University of Chinese Academy of Sciences, Beijing 100049, China}}

\newcommand{\sucas}{\affiliation{Southern Center for Nuclear-Science Theory (SCNT), Institute of Modern Physics,
Chinese Academy of Sciences, Huizhou 516000, Guangdong Province, China}}

\begin{document}

\title{ Hidden charmonium decays of spin-2 partner of $X(3872)$ }

\author{Yuanxin Zheng}\qfnu

\author{Zuxin Cai}\qfnu

\author{Gang Li}\email{gli@qfnu.edu.cn} \qfnu \itp

\author{Shidong Liu}\email{liusd@qfnu.edu.cn} \qfnu

\author{Jiajun Wu}\ucas\sucas%\email{wujiajun@ucas.ac.cn}

\author{Qi Wu}\email{wuqi@htu.edu.cn} \hnnu

\begin{abstract}
%It is generally expected that the $X(3872)$ has a spin-2 partner, $X_2$, whose quantum numbers are $J^{PC}=2^{++}$. In the hadronic molecular model, its mass is predicted to be below the $D^*\bar{D}^*$ threshold. And the new structure reported by the Belle Collaboration with mass $M = (4014.3 \pm 4.0 \pm 1.5)$ MeV and decay width $\omega = (4 \pm 11 \pm 6)$ MeV.
The Belle collaboration recently reported a promising candidate for the spin-2 $D^*\bar{D}^*$ partner of the $X(3872)$, called the $X_2$ for short, having a mass of $(4014.3 \pm 4.0 \pm 1.5)~\mathrm{MeV}$ and a width of $(4 \pm 11 \pm 6)~\mathrm{MeV} $. Assuming the $X_2$ as a pure molecule of the $D^*\bar{D}^*$, we calculated in detail the hidden charmonium decays of the $X_2 \to J/\psi V$ and $X_2\to\eta_cP$ via the intermediate meson loops, where $V = \rho^0\,,\omega$ and $P= \pi^0\,,\eta\,,\eta'$. The results indicate that the decay widths are strongly dependent on the $X_2$ mass. At present center value of the mass $4014.3~\mathrm{MeV}$, the width for the $X_2\to J/\psi\rho^0$ is predicted to be a few tens of keV, while it is on the order of $10^{2\text{-}3}~\mathrm{keV}$ for the $X_2\to J/ \psi\omega$; the predicted width for the $X_2\to \eta_c \pi^0$ is about a few keV, while the widths for $X_2\to\eta_c\eta$ and $\eta_c\eta'$ are around a few tens and tenths of keV, respectively. We also investigated the dependence of the ratios between these widths on the $X_2$ mass and on the $\eta$-$\eta'$ mixing angle, which may be good quantities for experiments. We hope that the present calculations would be checked experimentally in the future.

\end{abstract}

\date{\today}

%\pacs{14.40.Pq, 13.20.Gd, 12.39.Fe}

%14.40.Rt Exotic mesons

%13.75.Lb Meson-meson interactions

%13.20.Gd Decays of J/\psi, and other quarkonia

%14.40.Pq Heavy quarkonia

%14.40.Lb Charmed mesons

\maketitle

\section{Introduction}\label{sec:introduction}

Studies of exotic states received revolutionary developments in 2003 when Belle collaboration observed a structure in the $\pi^+\pi^-J/\psi$ invariant mass spectrum \cite{2003bellecollaborationPRL91-262001}, the $X(3872)$. Those candidates for exotic states are usually referred to collectively as \textit{XYZ} states. In order to understand the nature of the exotic states, numerous experimental and theoretical investigations have been performed and are ongoing or planned. Considerable effort is devoted to explanation of the \textit{XYZ} mass, width, and quantum numbers $J^{PC}$, which, in turn, helps us to understand their internal structures. There have been many reviews concerning this topic, e.g., the recent comprehensive ones \cite{2020PRbrambilla,2017lebedPiPaNP93-143,2016chenPR639-1} from experimental and theoretical status and perspectives.

Of the \textit{XYZ} states, the $X(3872)$ is the first and most well-studied representative. At present, the world average mass of the $X(3872)$ is $(3871.65\pm 0.06)~\mathrm{MeV}$ with a narrow full width $(1.19\pm 0.21)~\mathrm{MeV}$ \cite{2022pdgPoTaEP2022-083C01}. The $X(3872)$ quantum number $J^{PC}$ was eventually determined to be $1^{++}$ in 2013 by the LHCb experiments at CERN~\cite{2013lhcbcollaborationPRL110-222001,2015lhcbcollaborationPRD92-011102}. The $X(3872)$ lies extremely close to the $D^0 \bar{D}^{*0}$ threshold of $3871.69~\mathrm{MeV} $ and has a lager decay rate to $D^0 \bar{D}^{*0}$ \cite{2022pdgPoTaEP2022-083C01} so that it is naturally considered as a loosely bound mesonic molecule. The molecular model for the $X(3872)$ structure, therefore, becomes quite popular and successful to interpret the properties of the $X(3872)$ since its discovery \cite{2008flemingPRD78-094019,2011mehenPRD83-094009,2021mengPRD104-094003,2021wuEPJC81-193,2022wangPRD106-074015,2015guoPLB740-42,2015guoPLB742-394,2015mehenPRD92-034019,2013guoPLB725-127,2010dongJPGNPP38-015001,2009dongPRD79-094013,2008dongPRD77-094013,2004tornqvistx-}. Guo et al have given a comprehensive review on hadronic molecules \cite{2018guoRMP90-015004} and the especial review on the $X(3872)$ in the molecular model has been published recently \cite{2019kalashnikovaP62-568}.

Due to the great achievement of the molecular model, the $X(3872)$ has been often used a basis for predicting possible exotic states. Under the condition that the $X(3872)$ is a mesonic molecule of the $D\bar{D}^*$ with $J^{PC} = 1^{++}$, the heavy quark spin symmetry (HQSS) predicts existence of an isoscalar $2^{++}$ $D^*\bar{D}^*$ partner of the $X(3872)$~\cite{2012nievesPRD86-056004,2013hidalgo-duquePRD87-076006,2013hidalgo-duquePLB727-432,2013guoPRD88-054007,2016baruPLB763-20} (Following early papers~\cite{2015albaladejoEPJC75-547,2015guoPLB740-42,2023shiPLB843-137987}, we also call this partner state $X_2$ for short.). Its theoretically predicted mass is $m_{X_2} = 4012~\mathrm{MeV}$ with a narrow width on the same order as that of the $X(3872)$. Subsequently, lots of theoretical work partly or specially study the $X_2$ from different points of view~\cite{2013albaladejoPRD88-014510,2012sunCPC36-194,2013guoPRD88-054007,2013hidalgo-duquePLB727-432,2013hidalgo-duquePRD87-076006,2016baruPLB763-20,2021wangIJMPA36-2150107,2023shiPLB843-137987}. In particular, authors in Ref. \cite{2015albaladejoEPJC75-547} predict a small width of a few MeV based on an effective field theory. A very recent investigation studied the radiative decays $X_2\to \gamma \psi$ ($\psi = J/ \psi\,,\psi(2S)$) and indicated that the width ratio of the $X_2\to \gamma \psi(2S)$ to $X_2\to\gamma J/ \psi$ is smaller than unity, nearly equal to the corresponding one of the $X(3872)$~\cite{2023shiPLB843-137987}.

Experimental situation seems to achieve a possible breakthrough in 2022 when the Belle collaboration observed a structure in the invariant mass distribution of the $\gamma \psi(2S)$, which has a mass of $(4014.3\pm 4.0 \pm 1.5)~\mathrm{MeV}$ and a width of $(4\pm 11\pm 6) ~\mathrm{MeV}$ \cite{2022bellecollaborationPRD105-112011}. It is noted that the mass and width of the new structure agree well with those of the $X_2$, which, therefore, suggests it as a good candidate for the $X_2$ although the experiments gave a low global significance of $2.8\sigma$ \cite{2022bellecollaborationPRD105-112011}. It is believed that more and more experiments would follow up soon to provide more detailed information about the new structure, thereby contributing to its identification. The investigations related to this possible spin-2 $D^*\bar{D}^*$ exotic state are urgent, not only experimentally but also theoretically.

In this work, we systematically investigate the hidden charmonium decays of the $X_2 \to J/\psi V$ ($V = \rho^0\,,\omega$) and $X_2 \to \eta_c P$ ($P = \pi^0\,,\eta\,,\eta'$) in the molecular picture where the $X_2$ is assumed to be a pure mesonic molecule of the $D^*\bar{D}^*$ pair. Based on the effective field theory, we only considered in the calculations the contributions via the intermediate meson loops that have been widely used in the productions and decays of the exotic states (see, for example, Refs. \cite{2013guoPLB725-127,2008dongPRD77-094013,2009dongPRD79-094013,2010dongJPGNPP38-015001,2015guoPLB742-394,2023wuPRD107-034028,2023wux-,2023wuPRD107-054044,2023wangEPJC83-186,2023shiPLB843-137987,2023liEPJC83-258,2023jiaPRD107-074029,2022wangPRD106-074015,2022wangPRD106-074026,2022qianPLB833-137292,2022duPRD105-074018,2022caoEPJC82-161,2021wuEPJC81-193,2021wuPRD104-074011,2019wuPRD99-034022,2019guoPRL122-202002,2017linPRD95-114017,2016wuAHEP2016-e3729050,2016wuPRD94-014015,2016wangFP11-111402,2016guoPRD93-054009,2015mehenPRD92-034019,2015liPRD91-034020,2015guoPLB740-42,2015chenPoTaEP2015-043B05,2014liPRD90-054006,2013wangPRL111-132003,2013liPRD87-034020,2013liPRD88-094008}). Our basic concern here is to predict the partial decay widths of the processes mentioned above and give possible influence factors on the widths, such as the $X_2$ mass and the $\eta$-$ \eta'$ mixing angle. The calculated results show that the widths of the processes we considered depend strongly on the $X_2$ mass, indicating the importance of the precise mass of the $X_2$.

The rest of the paper is organized as follows. In Sec.~\ref{sec:formula},
we present the theoretical framework used in this work. Then in Sec.~\ref{sec:results} the numerical results are presented, and a brief summary is given in Sec.~\ref{sec:summary}.

\section{Theoretical Framework}\label{sec:formula}

\begin{figure}
    \centering
    \includegraphics[width=0.95\hsize]{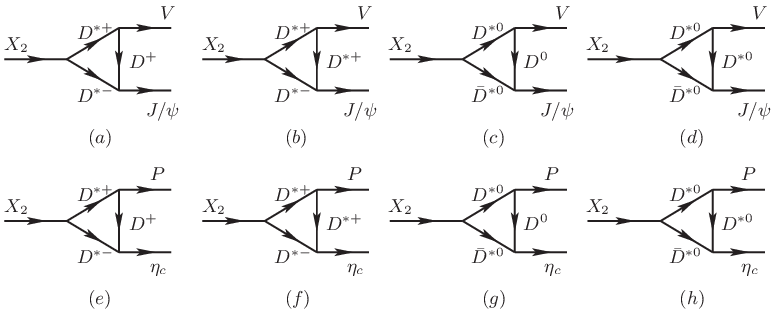}
    \caption{Feynman  diagrams for the processes $X_2 \to J/\psi V$ [(a)-(d)] and $X_2\to \eta_c P $ [(e)-(h)] with $V = \rho^0\,,\omega$ and $P = \pi^0\,,\eta\,,\eta'$ via charmed meson loops.}
    \label{fig:loops}
\end{figure}

\subsection{Effective interaction Lagrangians}
\label{sec:effective Lagrangians}
Figure \ref{fig:loops} shows the charmed meson loops devoted to the hidden charmonium decay processes of the $X_2$ that we consider in this work. The loop amplitudes for the decays $X_2\to J/\psi V$ and $X_2\to \eta_c P$ can be obtained by the sum of the corresponding diagrams.

Here we assume that the $X_2$ is an \textit{S}-wave molecular state with quantum numbers $I(J^{PC})=0(2^{++})$ given by the superposition of the $D^{*0}{\bar D}^{*0}$ and $D^{*+} D^{*-}$ hadronic configurations as
\begin{equation}\label{eq:WX2}
    |X_2 \rangle = \frac{1}{\sqrt{2}} \big( | D^{*0}{\bar D}^{*0} \rangle +  |D^{*+} D^{*-} \rangle\big) \,.
\end{equation}
%where $\theta$ is a phase angle describing the proportion of neutral and charged constituents.

Similar to the case for the $X(3872)$ \cite{2015guoPLB742-394}, the interaction of the $X_2$ with a pair of charmed and anticharmed mesons $D^*\bar{D}^*$ is described by \cite{2023shiPLB843-137987}
\begin{multline}\label{eq:LX2}
    \mathcal{L}_{X_2} = \frac{1}{\sqrt{2}}\big(\chi_\mathrm{nr}^0 X_{2\mu\nu} D^{*0\dagger \mu}\bar{D}^{*0\dagger\nu} \\+ \chi_{\mathrm{nr}}^\mathrm{c} X_{2\mu\nu} D^{*+\dagger \mu}D^{*-\dagger\nu}\big) + \mathrm{H.c.}\,.
\end{multline}
Here and later the symbols with (without) dagger index indicate the creation (annihilation) of the corresponding states. It is assumed that the $X_2$ is a pure $D^*\bar{D}^*$ bound state. In terms of such hadronic molecule picture, the coupling constants $\chi_{\mathrm{nr}}$'s are determined as \cite{2023shiPLB843-137987,2013guoPLB725-127,1965weinbergPR137-B672,2004baruPLB586-53}
\begin{equation}\label{eq:chinr}
    \chi_{\mathrm{nr}} = \left(  \frac{16\pi}{\mu}\sqrt{\frac{2E_\mathrm{B}}{\mu}}\right)^{1/2}\,,
\end{equation}
where $E_\mathrm{B}$ and $\mu$ are the binding energy of the $X_2$ relative to the $D^*\bar{D}^*$ threshold and the $D^*\bar{D}^*$ reduced mass, respectively. For the case of the $D^{*0}\bar{D}^{*0}$, $\chi_\mathrm{nr}^0$ is $1.32~\mathrm{GeV^{-1/2}}$, whereas it is $2.36~\mathrm{GeV^{-1/2}}$ for the $D^{*+}D^{*-}$. 

In the heavy quark limit, the interactions of the $S$-wave charmonia $J/\psi$ and $ \eta_c$ with the $D$ and $D^*$ mesons are described by the Lagrangian \cite{2004colangeloPRD69-054023}
\begin{eqnarray}\label{eq:Lpsi}
    	\mathcal{L}_{S} &=& \ii g_{\psi DD} \psi^{\dagger}_\mu \bar{D} \pararrowk{\mu}D \nonumber\\
    	&+& g_{\psi D^*D}\epsilon_{\mu\nu\alpha\beta}\partial^\mu \psi^{\dagger\nu} (D^{*\alpha} \pararrowk{\beta}\bar{D} - D \pararrowk{\beta}\bar{D}^{*\alpha}) \nonumber\\
    & -&\ii g_{\psi D^*D^*} \psi^{\dagger}_\mu (D^*_\nu \pararrowk{\nu}\bar{D}^{*\mu} + D^{*\mu}\pararrowk{\nu}\bar{D}^*_\nu \nonumber\\
    &-& D^*_\nu \pararrowk{\mu}\bar{D}^{*\nu}) \nonumber\\
    & -& g_{\eta_c D^*D^*}\epsilon_{\mu\nu\alpha\beta} \partial^\mu \eta_c^\dagger D^{*\nu}\pararrowk{\alpha}\bar{D}^{*\beta} \nonumber\\
    &+&\ii g_{\eta_c D^*D} \eta_c (D\pararrowk{\mu} \bar{D}^{*}_\mu + D_\mu^*\pararrowk{\mu}\bar{D} )+ \mathrm{H.c.}\,,
\end{eqnarray}
where $D=(D^0\,,D^+\,,D_s^+)$ and $D^*=(D^{*0}\,,D^{*+}\,,D_s^{*+})$ are the pseudoscalar and vector charmed meson triplet, respectively. The $\bar{D}^{(*)}$'s are the corresponding anticharmed meson triplets. In fact, the strange charmed mesons are not needed to be considered in present work thanks to no strange component in the $X_2$. The coupling constants $g_{\psi D^{(*)}D^{(*)}}$'s are related to the gauge coupling $g_1 = \sqrt{m_\psi}/(2m_D f_\psi)$ with the $J/\psi$ decay constant $f_\psi = 426~\mathrm{MeV}$ \cite{2004colangeloPRD69-054023,2021wuEPJC81-193}, namely
\begin{subequations}\label{eq:gpsidds}
\begin{align}
    g_{\psi DD} &= 2g_1 m_D \sqrt{m_\psi}\,,\label{eq:gpsidd}\\
    g_{\psi D^*D}&= 2g_1 \sqrt{m_\psi m_{D^*}/m_D}\,,\label{eq:gpsidsd}\\
    g_{\psi D^*D^*}& = 2g_1 m_{D^*}\sqrt{m_\psi}\,,\label{eq:gpsidsds}\\
    g_{\eta_c D^*D^*}& = 2g_1 \sqrt{m_{\eta_c}}\,,\label{eq:getacdsds}\\
    g_{\eta_cD^*D}& = 2g_1 \sqrt{m_{\eta_c} m_{D^*}m_D}\,.
\end{align}
\end{subequations}

Based on the heavy quark limit and chiral symmetry, the interactions between the light vector and pseudoscalar charmed mesons read \cite{2021wuEPJC81-193,2022wangPRD106-074015,2023wuPRD107-034028}
\begin{eqnarray}\label{eq:LV}
    	\mathcal{L} &=& -\ii g_{DDV} D_i^\dagger \pararrowk{\mu}D^{j}(V_\mu^\dagger)_j^i \nonumber\\
    	&-&2 f_{D^*DV}\epsilon_{\mu\nu\alpha\beta} (\partial^\mu V^{\nu\dagger})_j^i(D_i^\dagger \pararrowk{\alpha}D^{*\beta j} \nonumber\\
    	&-& D_i^{*\beta\dagger}\pararrowk{\alpha}D^{j})
    +\ii g_{D^*D^*V}D_i^{*\nu\dagger}\pararrowk{\mu}D_\nu^{*j}(V_\mu^\dagger)_j^i\nonumber \\
    &+& \ii 4 f_{D^*D^*V} D_{i\mu}^{*\dagger}(\partial^\mu V^{\nu\dagger} - \partial^\nu V^{\mu\dagger})_j^i D_\nu^{*j}\nonumber\\
    &-&\ii g_{D^*DP}\big(D^{i \dagger}\partial^{\mu} P_{ij}^\dagger D_\mu^{*j} - D_\mu^{*i\dagger}\partial^\mu P_{ij}^\dagger D^j\big) \nonumber\\
    &+& \frac{1}{2} g_{D^*D^*P}\epsilon_{\mu\nu\alpha\beta} D_i^{*\mu\dagger}\partial^\nu P^{ij\dagger}\pararrowk{\alpha} D_j^{*\beta} + \mathrm{H.c.}\,,
\end{eqnarray}
where the $V$ and $P$ are, respectively, the nonet vector and pseudoscalar mesons in the matrix form
\begin{subequations}\label{eq:vmatrix}
\begin{align}
    V &= 
    \begin{pmatrix}
    \frac{\rho^0}{\sqrt{2}}+\frac{\omega}{\sqrt{2}}&\rho^+&K^{*+}\\
    \rho^-&-\frac{\rho^0}{\sqrt{2}}+\frac{\omega}{\sqrt{2}}&K^{*0}\\
    K^{*-}&\bar{K}^{*0}&\phi
    \end{pmatrix}\,,\\
	P &= \begin{pmatrix}
		\frac{\pi^0}{\sqrt{2}} + \frac{\delta \eta + \gamma \eta'}{\sqrt{2}} & \pi^+ & K^+\\
		\pi^- & -\frac{\pi^0}{\sqrt{2}} + \frac{\delta \eta + \gamma \eta'}{\sqrt{2}} & K^0\\
		K^- & \bar{K}_0 & - \gamma \eta + \delta \eta' \label{eq:P}
	\end{pmatrix}\,.
\end{align}
\end{subequations}
Here $\delta = \cos(\theta_\mathrm{P} + \arctan\sqrt{2})$ and $\gamma = \sin(\theta_\mathrm{P} + \arctan\sqrt{2})$ with the $\eta$-$\eta'$ mixing angle $\theta_\mathrm{P}$ ranging from $-24.6^\circ$ to $-11.5^\circ$ \cite{2013liPRD88-014010,2023wuPRD107-034028,2012wangPRD85-074015,1996amslerPRD53-295,1983rosnerPRD27-1101,2022pdgPoTaEP2022-083C01}.
The coupling constants $g_{D^{(*)}D^{(*)}V}$'s and $g_{D^{(*)}D^{(*)}P}$'s could be determined using the following relations \cite{2021wuEPJC81-193}
\begin{subequations}\label{eq:gddvs}
\begin{align}
    g_{DDV} &= g_{D^*D^*V} = \frac{\beta g_V}{\sqrt{2}}\,,\label{eq:gddvgdsdsv}\\
    f_{D^*DV}&= \frac{f_{D^*D^*V}}{m_{D^*}} = \frac{\lambda g_V}{\sqrt{2}}\,,\label{eq:fdsdvfdsdsv}\\
    g_{D^*D^*P} &= \frac{g_{D^*DP}}{\sqrt{m_Dm_{D^*}}}= \frac{2g}{f_{\pi}}\,.\label{eq:gddp}
\end{align}
\end{subequations}
Here $\beta=0.9$ and $g_V = m_\rho / f_\pi$ with the pion decay constant $f_\pi = 132~\mathrm{MeV}$
\cite{1997casalbuoniPR281-145}. Moreover, $\lambda=0.56~\mathrm{GeV^{-1}}$ and $g = 0.59$ based on the matching of the form factors obtained from the light cone sum rule and from the lattice QCD calculations \cite{2003isolaPRD68-114001}.

It is recalled that the mass of the $X_2$ is close to the thresholds of the $D^{*}\bar{D}^{*}$. Consequently, the two charmed mesons $D^*$ and $\bar{D}^*$ interacting with the $X_2$ could be considered to be nearly on shell. However, another exchanged charmed meson in the loops is off shell. To account for the off-shell effect as well as the inner structure of the exchanged meson, a monopole form factor was included in the calculations~\cite{1994tornqvistZPC-PaF61-525,2005chengPRD71-014030}
\begin{equation}
    F(q^2,m^2) = \frac{m^2-\Lambda^2}{q^2-\Lambda^2}\,,
\end{equation}
where $q$ and $m$ are the momentum and mass of the exchanged meson, respectively; $\Lambda = m+\alpha \Lambda_\mathrm{QCD}$ with $\Lambda_\mathrm{QCD} = 0.22~\mathrm{GeV}$. The model parameter $\alpha$ could not be determined from the first principle, but its value was found to be of order of unity and depends not only on the exchanged particle but also on the external particles involved in the strong interaction~\cite{1994tornqvistZPC-PaF61-525,2005chengPRD71-014030}.  A search in the literature \cite{2022wangPRD106-074015,2020wangEPJC80-475,2017wuEPJC77-336,2016wuPRD94-014015,2006liuPRD74-074003,2009liuPLB675-441,2010chenPRD81-074006,2009zhangPRL102-172001,2011liPRD84-074005,2013liPRD88-014010,2016guoPRD93-054009} yields that for the decays of charmonium(-like) particles through the charmed meson loops the parameter $\alpha$ is commonly taken to be smaller than 2. In the present calculations, we vary $\alpha$ from 0.7 to 1.4 to exhibit its influence on the decay processes we considered.

\subsection{Amplitudes of $X_2\to J/\psi V$ and $X_2\to \eta_c P$ }
According to the Lagrangians above, the amplitudes $\mathcal{M}_V$ for the case $X_2\to J/\psi V$ ($V = \rho^0\,,\omega$) and $\mathcal{M}_P$ for $X_2\to \eta_c P$ ($P = \pi^0\,,\eta\,,\eta'$), governed by the loops in Fig.~\ref{fig:loops}, have the form
\begin{subequations}
	\begin{align}
		\mathcal{M}_V &= \frac{1}{2}  \chi_{\mathrm{nr}}^{c(0) }\sqrt{m_{X_2}}m_{D^*}\nonumber \\
		&\times \varepsilon^{\mu\nu}(X_2)\varepsilon^{*\alpha}(V)\varepsilon^{*\beta}(J/\psi) I_{\mu\nu\alpha\beta}\,,\\
		\mathcal{M}_P &= \frac{x}{2}  \chi_{\mathrm{nr}}^{c(0) }\sqrt{m_{X_2}}m_{D^*}\varepsilon^{\mu\nu}(X_2) I_{\mu\nu}\,,
	\end{align}
\end{subequations}
where the mass factor $\sqrt{m_{X_2}}m_{D^*}$ results from the non-relativistic normalization of the heavy fields involved in the $X_2$ vertex. $\varepsilon^{\mu\nu}(X_2)$, $\varepsilon^{*\alpha}(V)$, and $\varepsilon^{*\beta}(J/\psi) $ describe the polarization tensor of the initial $X_2$, the polarization vectors of the final light vector particle $\rho^0$ or $\omega$, and the heavy $J/\psi$, respectively. The parameter $x$ is equal to 1, $\beta$, and $\gamma$ for the $\pi^0$, $\eta$, and $\eta'$ emission, respectively (see Eq. \eqref{eq:P}). The tensor structures $I_{\mu\nu\alpha\beta}$ and $I_{\mu\nu}$ are expressed as
\begin{subequations}
	\begin{align}
	I_{\mu\nu\alpha\beta}^{a(c)} &= \int \frac{\mathrm{d}^4q}{(2\pi)^4}g_{\mu\rho}g_{\nu\sigma} [2f_{D^*DV}\epsilon_{\delta\alpha\eta\xi}p_3^\delta(p_1+q)^\eta]\nonumber\\
	&\times [g_{\psi D^*D}\epsilon_{\delta\beta\gamma\eta}p_4^\delta(p_2-q)^\eta]S^{\rho\xi}(p_1,m_{D^*})\nonumber\\
	&\times S^{\sigma\gamma}(p_2,m_{D^*}) S(q,m_D)F(q^2,m_D^2)\,,\\
	I_{\mu\nu\alpha\beta}^{b(d)} &= \int \frac{\mathrm{d}^4q}{(2\pi)^4}g_{\mu\rho}g_{\nu\sigma} [4f_{D^*D^*V}(p_{3,\eta} g_{\alpha\xi} - p_{3,\xi} g_{\eta\alpha})\nonumber\\
	& - g_{D^*D^*V}(p_1+q)_\alpha g_{\xi\eta}]g_{\psi D^* D^*}[(p_2+q)_\delta g_{\beta\gamma} \nonumber\\
	&+ (p_2+q)_\gamma g_{\beta \delta} - (p_2+q)_\beta g_{\gamma \delta}]S^{\rho\xi}(p_1,m_{D^*}) \nonumber\\
	&\times S^{\sigma\gamma}(p_2,m_{D^*}) S^{\delta\eta}(q,m_D)F(q^2,m_D^2)\,,\\
	I_{\mu\nu}^{e(g)}&= \int \frac{\mathrm{d}^4q}{(2\pi)^4}g_{\mu\rho}g_{\nu\sigma} [g_{D^*DP} p_{3,\xi}] [-g_{\eta_c D^*D} (p_2-q)_\gamma] \nonumber\\
	&\times S^{\rho\xi}(p_1\,,m_{D^*}) S^{\sigma\gamma}(p_2\,,m_{D^*})S(q\,,m_D)\,,\\
	I_{\mu\nu}^{f(h)}&= \int \frac{\mathrm{d}^4q}{(2\pi)^4}g_{\mu\rho}g_{\nu\sigma} [-g_{\eta_c D^*D^*} \epsilon_{\beta \eta \lambda\gamma}p_4^\beta (p_2-q)^\lambda]\nonumber\\
	&\times [\frac{1}{2}g_{D^*D^* P} \epsilon_{\delta\beta\lambda \xi}p_3^\beta(p_1+q)^\lambda]S^{\rho\xi}(p_1,m_{D^*}) \nonumber\\
	&\times S^{\sigma\gamma}(p_2,m_{D^*}) S^{\delta\eta}(q,m_D)F(q^2,m_D^2)\,.
\end{align}
\end{subequations}
Here $S$ and $S^{\mu\nu}$, respectively, represent the propagators for the charmed mesons $D$ and $D^*$ in the following form
\begin{subequations}
	\begin{align}
		S(q,m_D) &= \frac{1}{q^2-m_D^2+\ii \epsilon}\,,\\
		S^{\mu\nu}(q,m_{D^*}) &= \frac{-g^{\mu\nu} + q^\mu q^\nu/m_{D^*}^2}{q^2-m^2_{D^*}+\ii \epsilon}\,.
	\end{align}
\end{subequations}

The processes $X_2\to J/\psi \rho^0$ and $X_2\to \eta_c \pi^0$ break isospin symmetry so that their amplitudes are given by the difference between the neutral and charged meson loops: $\mathcal{M}^{(a/e)}+\mathcal{M}^{(b/f)}-\mathcal{M}^{(c/g)}-\mathcal{M}^{(d/h)} $. On the contrary, the other processes we considered follow isospin conservation. Hence, their amplitudes can be obtained by summing the contributions from the neutral and charged meson loops: $\mathcal{M}^{(a/e)}+\mathcal{M}^{(b/f)}+\mathcal{M}^{(c/g)}+\mathcal{M}^{(d/h)} $.

\begin{figure}
	\centering
	\includegraphics[width=0.9\linewidth]{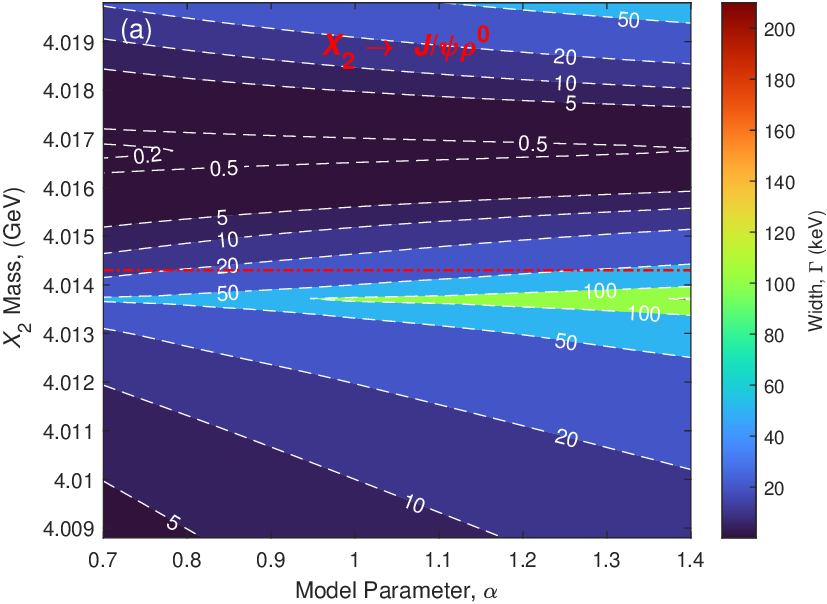}\\
	\includegraphics[width=0.9\linewidth]{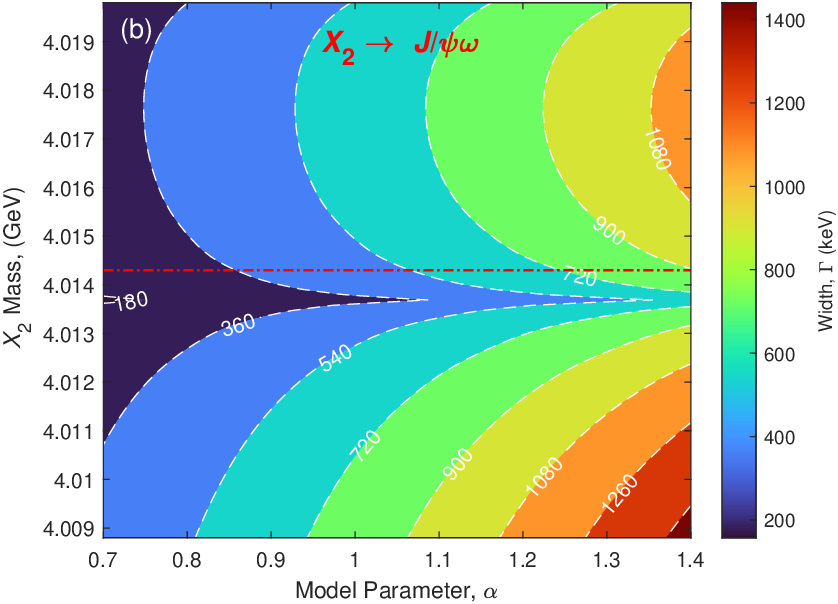}
	\caption{Widths of the decay processes $X_2\to J/\psi \rho^0$ (a) and $X_2\to J/\psi \omega $ (b) as a function of the $X_2$ mass $m_{X_2}$ and the model parameter $\alpha$. The white dashed lines with various numbers are the isolines of the widths. The red dash-dotted line represents the center mass 4.0143 GeV of the $X_2$.}
	\label{fig:x2tojpsirhoomg}
\end{figure}

\section{Numerical Results}\label{sec:results}

\subsection{Decay processes of $X_2\to J/\psi\rho^0 (\omega)$ }

In the following, we consider the decay widths of $X_2 \to J/\psi\rho^0$ and $J/\psi\omega$. In Fig.~\ref{fig:x2tojpsirhoomg} the decay widths for the both processes are shown for different $X_2$ masses and model parameters. We assume that the $X_2$ mass ranges from $4.009$ GeV to $4.020$ GeV. Moreover, the model parameter $\alpha$ is varied between 0.7 and 1.4. Within the intervals we considered, the width ranges from $\sim 0.2$ to $200~\mathrm{keV}$ for $X_2\to J/\psi \rho^0$, while for $X_2\to J/\psi \omega$ it is between 0.1 and 1.5 MeV. For a given $\alpha$ the width for $X_2\to J/\psi \rho^0$ first increases with the $X_2$ mass to a peak value at $m_{X_2} = 4.0137~\mathrm{GeV}$, then drops until $\sim 4.0167~\mathrm{GeV}$, and finally starts to increase again. However, the width for $X_2\to J/\psi \omega$ shows opposite variations with the $X_2$ mass at a given $\alpha$, i.e., the width exhibits a valley at $m_{X_2} = 4.0137~\mathrm{GeV} $, while near $m_{X_2} = 4.0175~\mathrm{GeV}$ it is a peak.

The opposite variation of the widths for these two processes with the $X_2$ mass can be understood in view of the fact that the process $X_2\to J/\psi \rho^0$ breaks isospin symmetry, while the process $X_2\to J/\psi \omega$ is of isospin conservation. In the molecular state scenario, the hidden charm decays of the $X_2$ occur via the charmed meson loops, where the interferences between the charged and neutral meson loops provide an important source of the isospin violation. Specifically, when $m_{X_2} = 4.0137~\mathrm{GeV}$, the coupling $\chi_{\mathrm{nr}}^0$ governing the interactions of the $X_2$ with the neutral $D^{*0}\bar{D}^{*0}$ pair becomes zero according to Eq. \eqref{eq:chinr}. Therefore, the destructive and constructive interference between the neutral and charged loops for the $X_2\to J/\psi \rho^0$ and $X_2\to J/\psi \omega$ both disappear. It, in turn, leads to a maximum width for $X_2\to J/\psi \rho^0$ at $m_{X_2} = 4.0137~\mathrm{GeV}$, but a minimum width for $X_2\to J/\psi \omega$.

However, near $m_{X_2} = 4.017~\mathrm{GeV}$ the contributions from the neutral and charged meson loops is close equal. If we ignore the mass difference of the neutral and charged charmed mesons, we could predict an approximate value of $m_{X_2} = 4.017~\mathrm{GeV}$ in terms of the following relation
\begin{subequations}\label{eq:semiana}
	\begin{align}
		\cos \Bigg[\arccos \Big(\frac{\chi_{\mathrm{nr}}^0}{\sqrt{(\chi_{\mathrm{nr}}^0)^2+(\chi_{\mathrm{nr}}^\mathrm{c})^2}}\Big)  + \theta \Bigg] &= 0\,,\\
		\sin \Bigg[\arcsin \Big(\frac{\chi_{\mathrm{nr}}^0}{\sqrt{(\chi_{\mathrm{nr}}^0)^2+(\chi_{\mathrm{nr}}^\mathrm{c})^2}}\Big) + \theta \Bigg] &= 1\,,
	\end{align}
\end{subequations}
where $\theta$ is a phase angle describing the proportion of neutral and charged constituents in $X_2$. In present work, we fixed $\theta$ to be $45^\circ$. Due to the mass difference of the neutral and charged mesons, the simple prediction deviates somewhat from the results in Fig. \ref{fig:x2tojpsirhoomg}. The inconsistency ($m_{X_2} = 4.0175~\mathrm{GeV}$ for $X_2\to J/ \psi \omega$ and $m_{X_2} = 4.0167$ for  $X_2\to J/ \psi \rho^0$) results from the different mass of the $\rho^0$ and $\omega$. These finding will be reproduced in the case of $X_2\to\eta_c P$ with $P = \pi^0$, $\eta$, and $\eta'$, which is discussed later.

It is seen that the widths for these two processes both increase as the model parameter $\alpha$ increases. Such $\alpha$-dependence could be canceled or weakened if we concentrate on the width ratio, namely
\begin{equation}\label{eq:Romg2rho}
	R_{\omega/\rho^0} = \frac{\Gamma (X_2\to J/ \psi \omega)}{\Gamma (X_2\to J/ \psi \rho^0)}\,. 
\end{equation}

\begin{figure}
	\centering
	\includegraphics[width=0.9\linewidth]{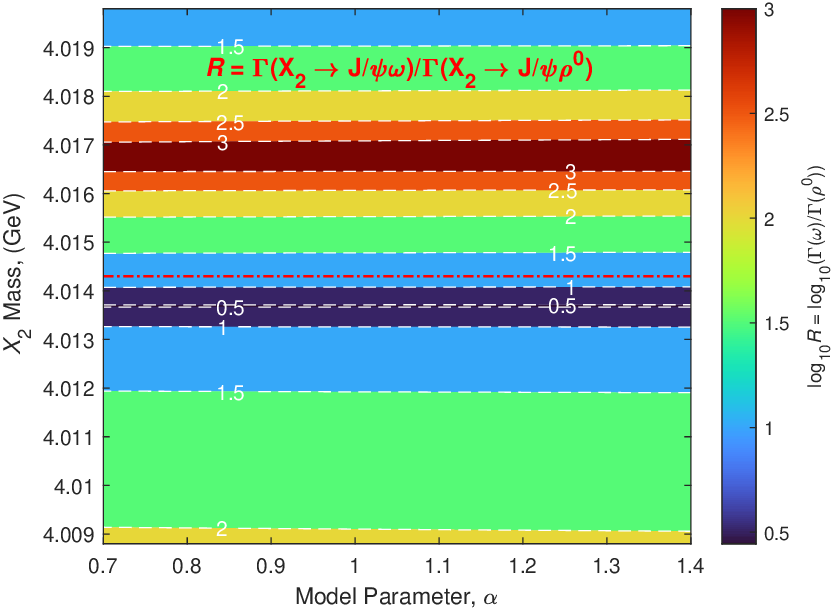}
	\caption{Ratios of the widths of the $X_2 \to J / \psi \omega$ to $X_2 \to J/ \psi \rho^0$. The white dashed lines with numbers represent the isolines of the logarithms of $R_{\omega/\rho^0}$ defined as Eq. \eqref{eq:Romg2rho}.}
	\label{fig:ratioo2r}
\end{figure}

We present the ratio $ R_{\omega/\rho^0}$ in Fig. \ref{fig:ratioo2r} as a function of the $X_2$ mass and the model parameter. It clearly shows that the ratio is rather independent of the model parameter $\alpha$. However, the $X_2$ mass dependence of the ratio show two extreme values at $m_{X_2} = 4.0137~\mathrm{GeV}$ and $m_{X_2} = 4.0167~\mathrm{GeV}$, as a result of the foregoing results in Fig.~\ref{fig:x2tojpsirhoomg}. Specially, at $m_{X_2} = 4.0137~\mathrm{GeV}$ where the width for $X_2\to J/ \psi\rho^0$ shows a peak value while it is minimum value for for $X_2\to J/ \psi \omega$, the ratio $R_{\omega/\rho^0}\simeq 3$; Near $m_{X_2} = 4.0167~\mathrm{GeV}$, the ratio $R_{\omega/\rho^0}$ is of the order of $10^3$. Moreover, at the center mass of the $X_2$, i.e., $m_{X_2} = 4.0143~\mathrm{GeV}$, the ratio $R_{\omega/ \rho^0}$ is about 15.

It is known from early experiments of BABAR \cite{2010thebabarcollaborationPRD82-011101}, Belle \cite{2005abex-}, and BESIII \cite{2019besiiicollaborationPRL122-232002} that the width ratio of the $X(3872)\to J/ \psi \pi^+\pi^-\pi^0$ to $X(3872)\to J/ \psi \pi^+\pi^-$ is around 0.8, 1.0, and 1.43, respectively. Consequently, using the world average branching fractions of $\omega\to \pi^+\pi^-\pi^0$ and $\rho^0 \to \pi^+ \pi^-$ \cite{2022pdgPoTaEP2022-083C01}, the ratio $\Gamma(X(3872) \to J/ \psi \omega) / \Gamma(X(3872) \to J/ \psi \rho^0)$ is roughly estimated to be 0.9, 1.1, and 1.60, respectively. If the $X_2$ decay modes are similar to those of the $X(3872)$, the width ratio $R_{\omega/\rho^0}$ for the $X_2$ should be about unity. To get such ratio, the $X_2$ has to be nearly pure $D^{*0}\bar{D}^{*0}$ or $D^{*+}D^{*-}$ molecular state according to our additional calculations that were performed under different phase angles (not shown here). For example, by reproducing the data of BESIII\cite{2019besiiicollaborationPRL122-232002}, the $D^{0}\bar{D}^{*0}$ component in $X(3872)$ is 
suggested to be (83--88\%) with the phase angle $\theta=66^\circ$--$70^\circ$~\cite{2021wuEPJC81-193}. However, in Ref. \cite{2009gamermannPRD80-014003} the ratio $R_{\omega/\rho^0} \simeq 1$ for the case of $X(3872)$ was explained by the larger effective phase space for the $J/\psi\rho^0$ decay than for $J/\psi \omega$, due to the large width of the $\rho^0$, which could compensate the suppression of the small mass difference between the neutral and charged charmed mesons. In the present case of the $X_2$, the influence of the $\rho^0$ width is found to be of minor importance because of the large mass difference between the $X_2$ and the $J/\psi$.

\subsection{Decay processes of $X_2 \to \eta_c\pi^0(\eta\,,\eta')$}

\begin{figure}
	\centering
	\includegraphics[width=0.9\linewidth]{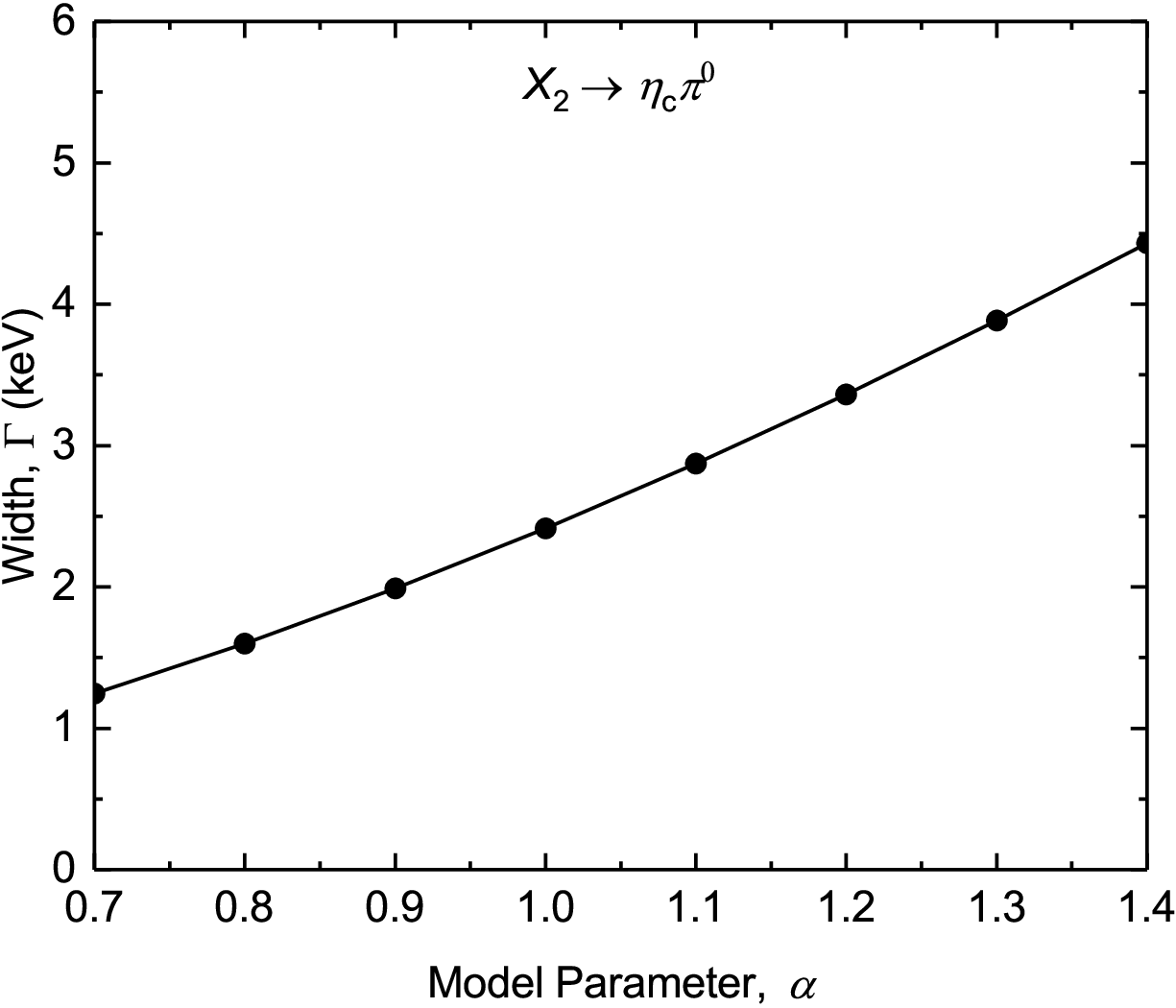}
	\caption{Widths of the $X_2\to \eta_c \pi^0$ as a function of the model parameter $\alpha$. The $X_2$ mass is taken to be $4.0143~\mathrm{GeV}$. The solid line is merely to guide the eye.}
	\label{fig:x2etacpi0}
\end{figure}

In this section we present the widths of the $X_2$ decaying to $\eta_cP$ ($P = \pi^0\,,\eta\,,\eta'$). Figure \ref{fig:x2etacpi0} show the width of the process $X_2\to \eta_c \pi^0 $. The width increases with increasing the model parameter, i.e., for $\alpha = 0.7$ the width amounts to 1.24 keV, which rises by a factor of about 4 to 4.43 keV at $\alpha = 1.4$.

\begin{figure}
	\centering
	\includegraphics[width=0.9\linewidth]{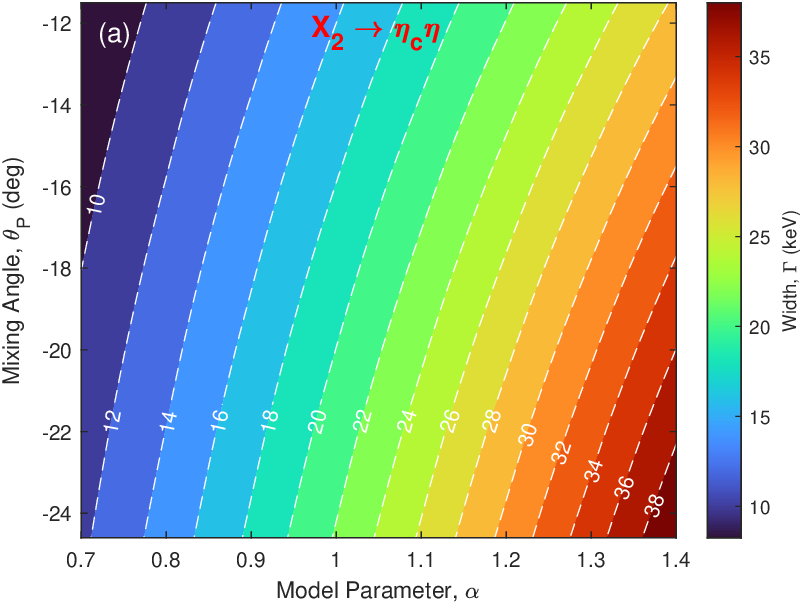}\\
	~\includegraphics[width=0.92\linewidth]{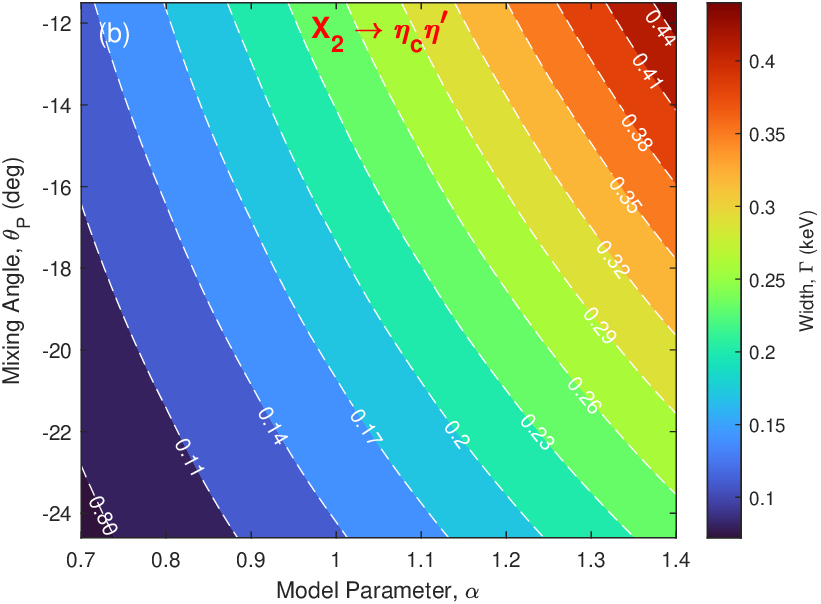}
	\caption{Widths for the $X_2\to \eta_c\eta$ (a) and $X_2\to \eta_c\eta'$ (b) as a function of the mixing angle $\theta_\mathrm{P}$ and the model parameter $\alpha$. The white dashed lines with various numbers are the isolines of the widths. The $X_2$ mass is taken to be $4.0143~\mathrm{GeV}$.}
	\label{fig:x2etaceta}
\end{figure}

For the processes $X_2\to\eta_c\eta$ and $X_2\to\eta_c\eta'$, the influence of the $\eta$-$\eta'$ mixing angle was considered in our calculations. Figure \ref{fig:x2etaceta} shows the widths of these two processes $X_2\to \eta_c\eta$ and $X_2\to \eta_c\eta'$ for different $\eta$-$\eta'$ mixing angles and model parameters. It is seen that these two widths both increase with increasing the model parameter $\alpha$ for a given mixing angle $\theta_\mathrm{P}$, similar to the foregoing cases of the $X_2\to \eta_c \pi^0$ and $X_2\to J/\psi\rho^0(\omega)$. However, their mixing angle dependence is different, i.e., at a given model parameter the width for $X_2\to \eta_c\eta$ decreases with increasing the mixing angle, while it increases for $X_2\to \eta_c\eta'$. The reason is as follows: The processes $X_2\to \eta_c\eta$ and $X_2\to \eta_c\eta'$ depend on the $\beta$ and $\gamma$, respectively. The $\gamma$ is increased with the increase of $\theta_P$ while $\beta$ is decreased. As a result, the decay width for the $X_2\to \eta_c\eta$ decreases with increasing the mixing angle while it is just the opposite for the case of the $X_2\to \eta_c\eta'$. It is worth noting that the variation of the widths with the mixing angle is not drastic, as they should.

For the $X_2\to \eta_c\eta$, it is seen from Fig. \ref{fig:x2etaceta}(a) that the width ranges from $\sim 8$ keV to 39 keV in the intervals of the mixing angle and model parameter that we consider. However, the width for $X_2\to \eta_c\eta'$ is just between $\sim 0.08$ keV and  $0.44$ keV (see Fig. \ref{fig:x2etaceta}(b)), two (one) orders of magnitude smaller than that for $\eta (\pi^0)$ decay.

% Similar to treatment for the case of the $X_2\to J/ \psi V$ ($V = \rho^0\,,\omega$), we define two ratios of the widths of the $X_2\to \eta_c\eta$ and $X_2\to\eta_c\eta'$ to that of the $X_2\to \eta_c \pi^0$, namely
% \begin{subequations}\label{eq:ratio}
% 	\begin{align}
% 		R_{\eta/\pi^0} &= \frac{\Gamma (X_2\to\eta_c\eta)}{\Gamma (X_2\to\eta_c\pi^0)}\,, \label{eq:Reta2pi0}\\
% 		R_{\eta'/\pi^0} &= \frac{\Gamma (X_2\to \eta_c\eta')}{\Gamma (X_2\to \eta_c\pi^0)}\,.\label{eq:Retap2pi0}
% 	\end{align}
% \end{subequations}
% The ratios obtained using the results in Figs. \ref{fig:x2etacpi0} and \ref{fig:x2etaceta} are shown in Fig.~\ref{fig:eta2pi}. As seen, the ratio $R_{\eta/\pi^0}$ is on the order of 10, while the ratio $R_{\eta'/\pi^0}$ is much smaller, of order $10^{-1}$. Strictly speaking, at a given mixing angle, these two ratios exhibit variation with the model parameter. However, the variation is seen to be actually small.

% \begin{figure}
% 	\centering
% 	\includegraphics[width = 0.9\linewidth]{eta2pi}\\
% 	~~\includegraphics[width = 0.92\linewidth]{etap2pi}
% 	\caption{Ratios of the widths of the $X_2\to \eta_c\eta$ (a) and $X_2\to\eta_c\eta'$ (b) to that of the $X_2\to \eta_c\pi^0$. The white dashed lines with various numbers are the isolines of the ratios obtained by Eq. \eqref{eq:ratio}. The $X_2$ mass is taken to be $4.0143~\mathrm{GeV}$.}
% 	\label{fig:eta2pi}
% \end{figure}

\begin{figure}
	\centering
	\includegraphics[width=0.9\linewidth]{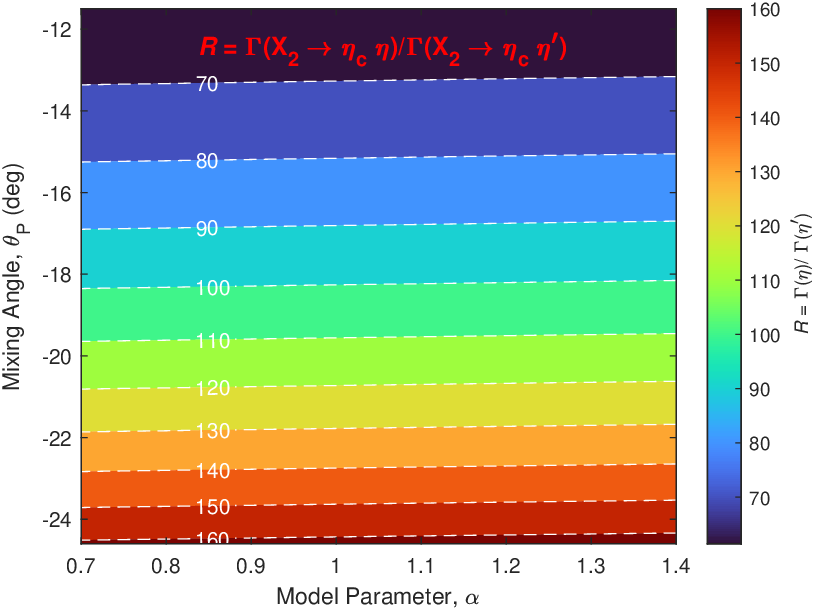}
	\caption{Ratios of the widths of the $X_2\to \eta_c\eta$ to $X_2\to \eta_c\eta'$. The white dashed lines with numbers represent the isolines of the ratio $R_{\eta/\eta'}$. The $X_2$ mass is taken to be $4.0143~\mathrm{GeV}$.}
	\label{fig:ratioeta2etap}
\end{figure}

Similar to treatment for the case of the $X_2\to J/ \psi V$ ($V = \rho^0\,,\omega$), we define the ratio of the widths of the $X_2\to \eta_c\eta$ to $X_2\to\eta_c\eta'$, namely
	\begin{align}\label{eq:ratio}
		R_{\eta/\eta'} &= \frac{\Gamma (X_2\to\eta_c\eta)}{\Gamma (X_2\to\eta_c\eta')}\tag{15}
	\end{align}
The ratio obtained using the results in Fig. \ref{fig:x2etaceta} is shown in Fig.~\ref{fig:ratioeta2etap}. 
%In Fig. \ref{fig:ratioeta2etap} we show the width ratio $R_{\eta/\eta'}$, defined as $\Gamma(X_2\to\eta_c\eta)/\Gamma(X_2\to\eta_c\eta')$. 
It is clearly seen that this ratio $R_{\eta/\eta'}$ is of order $10^2$ and quite independent of the model parameter. However, it decreases as the mixing angle $\theta_\mathrm{P}$ increases, i.e., at $\theta_\mathrm{P}=-24.6^\circ$ the ratio $R_{\eta/\eta'} \simeq 160 $, while it decreases at $\theta_\mathrm{P}=-11.5^\circ$ to $\sim 60$. The ratio $R_{\eta/\eta'}$ is expected to be measured in future, e.g. BESIII and Belle, which may help us constrain this mixing angle.

In the following, we focus on the $X_2$ mass dependence of the width for the decays $X_2\to\eta_cP$ ($P = \pi^0\,,\eta\,,\eta'$). Since the foregoing results indicate that the width ratio $R$'s are (nearly) of model parameter-$\alpha$ independence. Hence, we fixed $\alpha = 1$ in the following calculations. Again, the $X_2$ mass is assumed to range from $4.009$ GeV to $4.020$ GeV.

\begin{figure}
	\centering
	\includegraphics[width=0.9\linewidth]{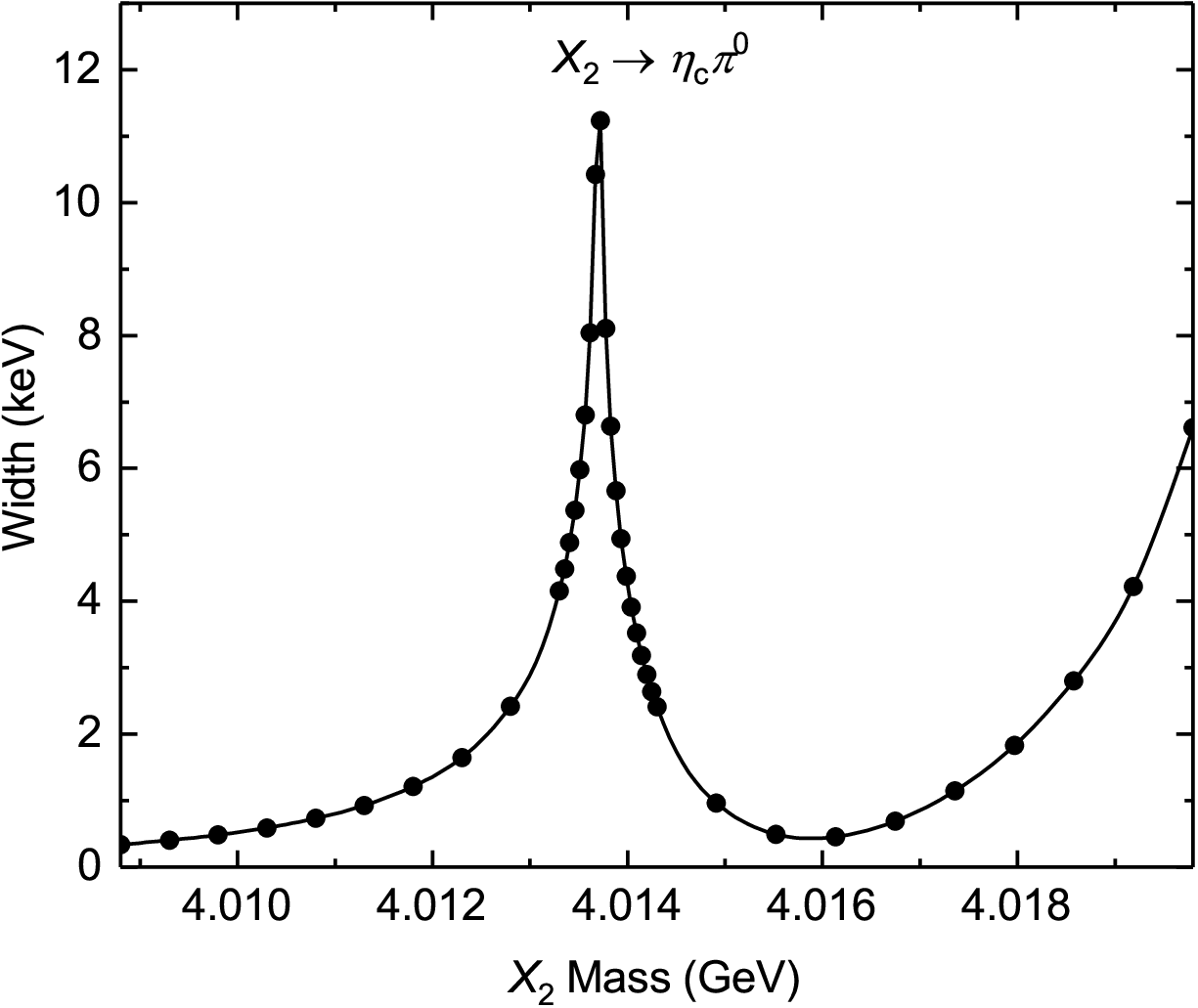}
	\caption{Widths of the $X_2\to \eta_c \pi^0$ for different $X_2$ masses. The model parameter $\alpha$ was fixed to be 1.}
	\label{fig:pi0mrun}
\end{figure}

\begin{figure}
	\centering
	\includegraphics[width=0.92\linewidth]{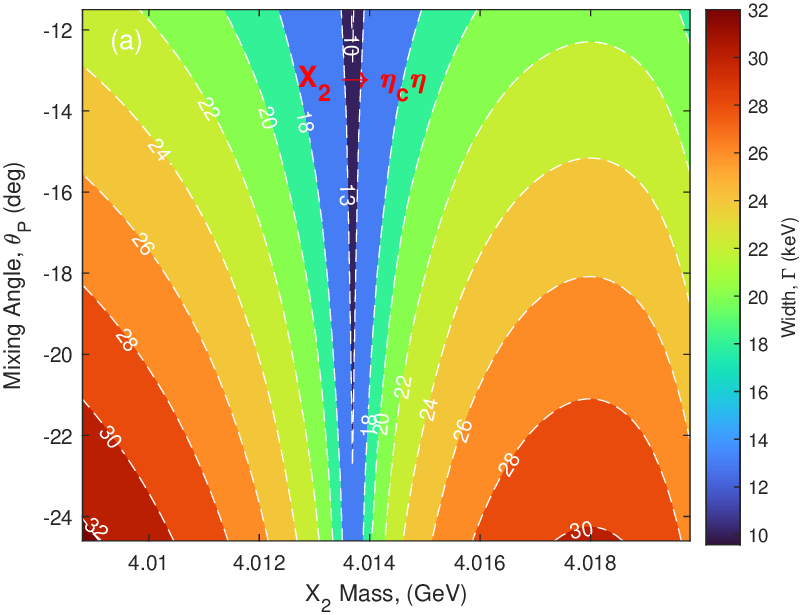}\\
	~\includegraphics[width=0.92\linewidth]{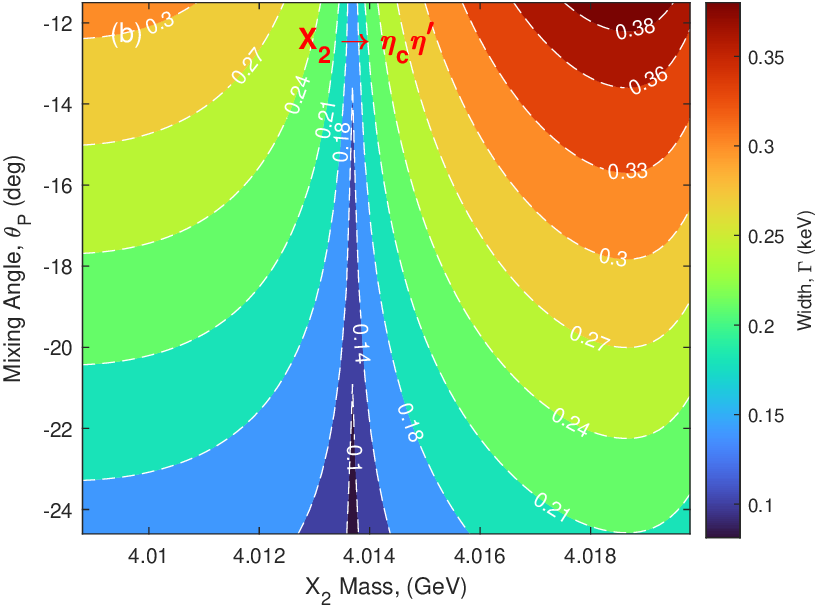}
	\caption{Widths for the $X_2\to \eta_c\eta$ and $X_2\to \eta_c\eta'$ as a function of the $\eta$-$\eta'$ mixing angle $\theta_\mathrm{P}$ and the $X_2$ mass. The model parameter $\alpha =1$.}
	\label{fig:etamrun}
\end{figure}

In Fig. \ref{fig:pi0mrun}, the width of the process $X_2\to\eta_c\pi^0$ is given for different $X_2$ masses. Figure \ref{fig:etamrun} shows the widths of the two processes $X_2\to \eta_c\eta$ and $X_2\to \eta_c\eta'$ for different $\eta$-$\eta'$ mixing angle $\theta_\mathrm{P}$'s and $X_2$ masses. The outstanding result of Figs. \ref{fig:pi0mrun} and \ref{fig:etamrun} is that the widths exhibit extreme values, similar to those in Fig. \ref{fig:x2tojpsirhoomg}. At $m_{X_2}= 4.0137~\mathrm{GeV}$, the width for $X_2\to \eta_c\pi^0$ exhibits a maximum value of about $12~\mathrm{keV}$, whereas the two decays of the $X_2$ to $\eta_c\eta$ and to $\eta_c\eta' $ show minimum widths. 

The extreme value of the widths at $m_{X_2}=4.0137~\mathrm{GeV}$ is straightforward in view of the fact that the $D^{*0}\bar{D}^{*0}$ threshold is $4.0137~\mathrm{GeV}$, thereby causing the coupling constant of the $X_2$ to the $D^{*0}\bar{D}^{*0}$ pair to be $\chi_{\mathrm{nr}}^0=0$ according to Eq. \eqref{eq:chinr}. Moreover, the decay process $X_2\to \eta_c\pi^0$ violates isospin symmetry, while the other decays of the $X_2$ to $\eta_c\eta$ and $\eta_c\eta' $ follow isospin conservation. As a result, the widths for all the processes we considered exhibit a extreme value at $m_{X_2} = 4.0137~\mathrm{GeV}$.

We can estimate the other extreme positions that are located between $m_{X_2} = 4.15$--$4.19~\mathrm{GeV}$, as shown in Figs. \ref{fig:pi0mrun} and \ref{fig:etamrun}, using a semi-quantitative analysis described as Eq. \eqref{eq:semiana}. It also yields the extreme point at $m_{X_2}\simeq 4.017~\mathrm{GeV}$, being the same with the case of $X_2\to J/\psi\rho^0(\omega)$. The deviation of the estimation from the results shown in Figs. \ref{fig:pi0mrun} and \ref{fig:etamrun} is mainly due to the different masses of the light mesons in the final states as well as the mass difference of the neutral and charged charmed mesons.%, namely the $u$ and $d$ quark mass difference.

% In Fig. \ref{fig:eta2pi0mrun}, the width ratios, namely, the widths of the $X_2\to \eta_c \eta$ and $X_2\to\eta_c\eta'$ to that of the $X_2\to \eta_c\pi^0$ defined explicitly as Eq. \eqref{eq:ratio}, are given for different $X_2$ masses and $\eta$-$\eta'$ mixing angles. As implied by the widths shown in Figs. \ref{fig:pi0mrun} and \ref{fig:etamrun}, at $m_{X_2} = 4.0137~\mathrm{GeV}$ the ratios $R_{\eta/\pi^0}$ and $R_{\eta'/\pi^0}$ show minimum values on the order of 1 and $10^{-2}$, respectively. On the contrary, these two ratios exhibit a peak near $m_{X_2} = 4.016~\mathrm{GeV}$. The variation of the ratios with the $X_2$ mass can be understood based on the widths shown in Figs.~\ref{fig:pi0mrun} and \ref{fig:etamrun}.

% \begin{figure}
% 	\centering
% 	\includegraphics[width=0.9\linewidth]{eta2pi0mrun}\\
% 	\includegraphics[width=0.9\linewidth]{etap2pi0mrun}
% 	\caption{Ratios of the widths of the $X_2\to \eta_c\eta$ (a) and $X_2\to\eta_c\eta'$ (b) to that of the $X_2\to \eta_c\pi^0$ as a function of the$\eta$-$\eta'$ mixing angle $\theta_\mathrm{P}$ and the $X_2$ mass $m_{X_2}$. The white dashed lines with various numbers are the isolines of the ratios obtained by Eq. \eqref{eq:ratio}.}
% 	\label{fig:eta2pi0mrun}
% \end{figure}

\begin{figure}
	\centering
	\includegraphics[width=0.92\linewidth]{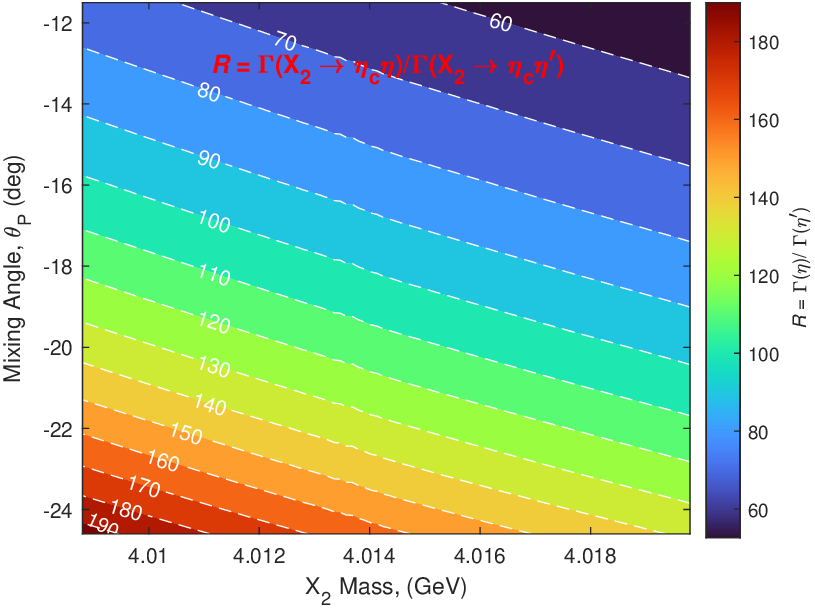}
	\caption{Ratios of the widths of the $X_2\to \eta_c\eta$ to $X_2\to \eta\eta'$ as a function of the $\eta$-$\eta'$ mixing angle and the $X_2$ mass. The white dashed lines with numbers represent the isolines of the ratio defined as $\Gamma(X_2\to\eta_c\eta)/\Gamma(X_2\to\eta_c\eta')$.}
	\label{fig:eta2etapmrun}
\end{figure}

Figure \ref{fig:eta2etapmrun} depicts the width ratio $R_{\eta/\eta'}$ for different $X_2$ masses, defined in Eq. \eqref{eq:ratio}. It is shown that this ratio decreases monotonously with increasing the $X_2$ mass and the $\eta$-$\eta'$ mixing angle. The extreme features existing in the widths (see Fig. \ref{fig:etamrun}) are canceled due to the fact that the widths for the $X_2\to\eta_c\eta$ and $X_2\to\eta_c\eta'$ have similar variation tendency with the $X_2$ mass\footnote{These two widths both first decrease with the $X_2$ mass to a minimum value at 4.0137 GeV, then increase to a peak value near 4.017 GeV, and finally decrease.}. The physics behind is that these two processes both remain the isospin symmetry.

\section{Summary}\label{sec:summary}

In this work, we investigated in detail the widths of the $X_2$ decaying to $J/ \psi V$ ($V = \rho^0\,,\omega$) and to $\eta_cP$ ($P = \pi^0\,,\eta\,,\eta'$) using the effective Lagrangian approach. In calculations, we assume the $X_2$ as a molecular state of the $D^{*0}\bar{D}^{*0} $ and $D^{*+}D^{*-}$ with equal proportion. Moreover, we only consider the contributions from the triangle hadron loops made of the charmed mesons $D^{(*)}$ and $\bar{D}^*$.

The calculated results indicate that the widths are all model-$\alpha$ dependent. However, the relative ratios between the widths of different processes are nearly model-$\alpha$ independent. It is found that the decays we considered are quite sensitive to the $X_2$ mass. This finding is straightforward since we considered the $X_2$ within the framework of the molecular picture, in which the coupling strength of the $X_2$ to the $D^*\bar{D}^{*}$ depends directly on the binding energy and, in turn, on the $X_2$ mass. In particular, near $m_{X_2} = 4.0137~\mathrm{GeV}$ and $4.017~\mathrm{GeV}$, the coupling strength of the $X_2$ to the neutral $D^{*0}\bar{D}^{*0}$ approaches zero and equals to the coupling strength between the $X_2$ and $D^{*+}D^{*-}$ pair, respectively. Consequently, at $m_{X_2} = 4.0137~\mathrm{GeV}$, the decays that violate the isospin symmetry exhibit widths of peak values, whereas those remaining the isospin symmetry show minimum widths. Near $m_{X_2}=4.017~\mathrm{GeV}$, the opposite happens. Accordingly, it indicate the significance of the precise measurement of the $X_2$ mass.

At the present $X_2$ center mass $m_{X_2}=4.0143~\mathrm{GeV}$, the width for the $X_2\to J/\psi \rho^0$ is a few tens of keV, while it is on the order of $10^{2-3}$ keV for the $X_2\to J/\psi \omega$. The corresponding width ratio $R_{\omega/\rho^0}$ is calculated to be about 15, one order of magnitude larger than that for the case of $X(3872)$ which approaches unity. 

For the other case of the $X_2\to\eta_cP$ ($P = \pi^0\,,\eta\,,\eta'$), we additionally considered the $\eta$-$\eta'$ mixing angle in the calculations. It is shown that the variation of decay widths due to the mixing angle is moderate, $\lesssim 2$ times change when mixing angle increases from $-24.6^\circ$ and $-11.5^\circ$. At the $X_2$ center mass $m_{X_2}=4.0143~\mathrm{GeV}$, the width for the $X_2\to \eta_c \pi^0$ is about a few keV, while the widths for $X_2\to\eta_c\eta$ and $\eta_c\eta'$ are around a few tens and tenths of keV, respectively.

\begin{acknowledgements}\label{sec:acknowledgements}

This work is partly supported by the National Natural Science Foundation of China under Grant Nos.
12075133, 12105153, 12175239, 12221005 and 11835015, and by the National Key R$\&$D Program of China under Contract No. 2020YFA0406400, and by the Chinese Academy of Sciences under Grant No. YSBR-101, and by the Natural Science
Foundation of Shandong Province under Grant Nos. ZR2021MA082, and ZR2022ZD26. It is also supported by Taishan
Scholar Project of Shandong Province (Grant No.tsqn202103062),
the Higher Educational Youth Innovation Science and Technology
Program Shandong Province (Grant No. 2020KJJ004).

\end{acknowledgements}

% References using bib
\bibliography{particlePhys.bib}
\end{document}